# DESIGN OF CONFORMAL ANTENNA FOR AIRCRAFT APPLICATIONS

*N.V.Laxminarayen[1*], G.Lokeshwaran[1*], K.S.Murugan[1*] and Dr.P.Jothilakshmi[1#]*

[1*]Final Year UG Students, [1#]Professor
Sri Venkateswara College of Engineering, Pennalur, Sriperumbudur
mail id: nvlnarayen96@gmail.com, jothi@svce.ac.in

*Abstract*— *The most challenging thing in real world is communicating with aircraft, even though several communication technologies has been adopted for tracking and monitoring the aircraft there is no cent percent efficiency, for that an implementation of conformal antenna for transmission and reception of the signal is preferred. In radio communication and avionics a conformal antenna or conformal array is a flat radio antenna which is designed to conform or follow some prescribed shape, for example a curved conformal antenna is designed and is mounted on or embedded in a curved surface. The conformal antenna is a collection of large number of smaller antennas (PAA) each one is connected to a phase shifter. The phased array antenna will have high directivity in desired application. Conformal arrays are typically limited to high frequencies in the UHF or microwave range, where the wavelength of the waves is small enough that small antennas can be used. The main objective of this project is to embed this conformal antenna on the surface of the aircraft with increased gain. To implement the above mentioned problem we are using CST microwave software. Conventional methods now used in aircrafts are done by using conformal antenna to save space and even military applications to be anonymous. Antenna stands to be an interface between the transmitter and the receiver. By working on the software and hardware features of antennas we can develop an antenna with better gain. By the adoption of better gain the antenna will be more efficient. Thus by implementing this conformal antenna on the aircraft surface with increased gain high degree of accuracy, clarity and effective communication link can be achieved.*

*Index Terms— Conformal antenna, Conformal array, PAA, UHF, CST microwave studio.*

## I. INTRODUCTION

HIS Antennas are key components of any wireless communication system. They are the devices that allow for the transfer of a signal to waves that propagates through space and can be received by another antenna. The receiving antennas responsible for the reciprocal process which is turning an electromagnetic wave into a signal or voltage at its terminals that can subsequently be processed by the receiver. The receiving and transmitting functionalities of the antenna structure itself are fully characterized by maxwell's equations.

An antenna system is defined as the combination of the antenna and its feed line. As an antenna is usually connected to a transmission line, how to best make this connection is a subject of interest, since the signal from the feedline should be radiated into space in an efficient and desired way. In some applications where space is very limited such as hand portables and aircraft, it is desirable to integrate the antenna and its feedline. In other applications such as the reception of TV broadcasting, the antenna is far away from the receiver and a long transmission line has to be used.

The dipole antenna, a straight wire, fed auto the center by a two-wire transmission line was the first antenna ever used and is also one of the best understood. For effective reception and transmission it must be approximately l/2 long auto the frequency of operation. It must be fairly long when used at low frequencies, and even auto higher frequencies its protruding nature makes it quite understandable. Further, its low gain, lack of directionality and extremely narrow bandwidth make it even less attractive.

Requirements for conformal antennas for airborne systems, increased bandwidth requirements, and multi functionality have led to heavy exploitation of printed (patch) or other slot-type antennas and the use of powerful computational tools for designing such antenna. Needless to say, the commercial mobile communications industry has been the catalyst for the recent explosive growth in antenna design needs. Certainly, the past decade has seen an extensive use of antennas by public for cellular, GPS, satellite, wireless LAN for computers Wi-Fi, Bluetooth technology, radio frequency ID devices, WiMAX and so on. However, future needs will be even greater when a multitude of antennas are integrated into automobiles for all sorts of communication needs.

Microstrip antennas (often called patch antennas) are widely used in the microwave frequency region because of their simplicity and compatibility with printed-circuit technology, making them easy to manufacture either as stand-alone elements or as elements of arrays. A microstrip antenna consists of a planar radiating structure over a ground plane separated by an electrically thin layer of dielectric substrate . The patch conductors can assume virtually any shape but conventional shapes are generally used to simplify analysis and performance prediction. The rectangular and circular patches are the basic and most commonly used microstrip antennas. These patches can be used for the simplest and the most demanding applications. For example, characteristics such as dual and circular polarizations, dual-frequency operation, frequency agility, broad bandwidth, feedline flexibility, beam scanning, omnidirectional patterning and soon are easily obtained in these patch shapes. Because these geometries are separable in nature, their analysis is also relatively simple.

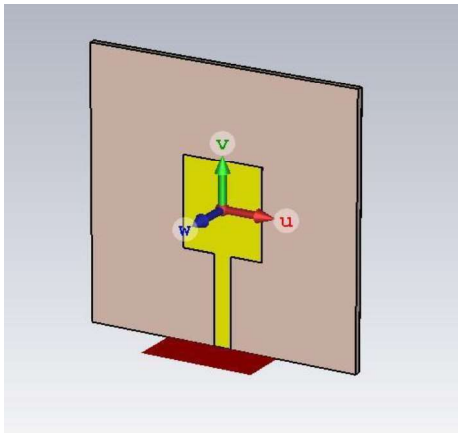

Fig 1.1 Microstrip Patch Antenna

## II. DESIGN OF CONFORMAL ANTENNA
*A. Overview*

Conformal antennas usually form a determined shape in order to accomplish certain aerodynamic or electromagnetic considerations, such as special angular coverage. Either omnidirectional or directional patterns with an arbitrary direction can be obtained using cylindrical, conical, or spherical radiating structures. Typically, the omnidirectional pattern can also be accomplished by means of conventional wire antennas such as monopoles or dipoles. Nevertheless, a strict tolerance manufacturing process is required for these wire antennas in very high frequency applications . In addition, the possibility to generate beam forming in conformal array antennas can provide more flexibility to the communication system. On the other hand, the micro strip conformal antennas are widely used in satellite, aerospace, or point-to-point communications due to many advantages such as low profile, low cost, or easy fabrication. Nevertheless, the power handling capacity of these kinds of antennas is limited, and the dielectric substrate losses as much as the bending of the structure can produce a significant efficiency reduction, especially for a large antenna array.

In this way, the conventional waveguide technology is suitable to resolve these issues due to its advantages of low losses, high power capacity, and high efficiency. In addition, a complex conformal structure can be approximated by using another shape with planar faces like a regular polyhedron, reducing the costs derived from the conformal structure manufactured using conventional waveguide technology.

The moving missile is required to realize the data transmission between the missile and the launcher for high precision attack. It leads to how to effectively design the missile antennas when considering the least disturbance to the flight dynamics and to the air frame structure. It should be with a low profile and can be mounted on the surface of a missile. Obviously, the cylindrically conformal antenna is a preferred choice. Such kind of antennas has always been the focus of the research and several forms of practical antennas were reported, such as printed dipole , microstrip patch, printed slot, and the substrate integrated waveguide (SIW) antennas.

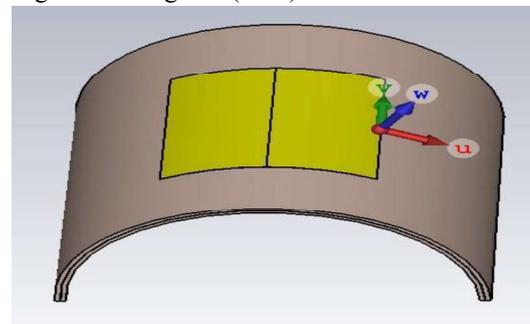

Fig 1.2 Conformal structure with single Microstrip Patch

*B. Design Equations*

First, the length of the patch L controls the resonant frequency as seen here. This is true in general, even for more complicated microstrip antennas that weave around the length of the longest path on the microstrip controls the lowest frequency of operation. Equation (5.1) below gives the relationship between the resonant frequency and the patch length:

$$f_c = \frac{1}{2L\sqrt{\varepsilon_0 \varepsilon_r \mu_0}}$$

The width W controls the input impedance and the radiation pattern (see the radiation equations here). The wider the patch becomes the lower the input impedance is. The permittivity of the substrate controls the fringing fields - lower permittivities have wider fringes and therefore better radiation. Decreasing the permittivity also increases the antenna's bandwidth. The efficiency is also increased with a lower value for the permittivity. The impedance of the antenna increases with higher permittivity.

Higher values of permittivity allow a "shrinking" of the patch antenna. Particularly in cell phones, the designers are given very little space and want the antenna to be a half- wavelength long. One technique is to use a substrate with a very high permittivity. Equation (5.2) above can be solved for L to illustrate this:

$$L = \frac{1}{2\sqrt{\varepsilon_0 \varepsilon_r \mu_0} f_c}$$

Hence, if the permittivity is increased by a factor of 4, the length required decreases by a factor of 2. Using higher values for permittivity is frequently exploited in antenna miniaturization.

The height of the substrate h also controls the bandwidth increasing the height increases the bandwidth. The fact that increasing the height of a patch antenna increases its bandwidth can be understood by recalling the general rule that "an antenna occupying more space in a spherical volume will have a wider bandwidth". This is the same principle that applies when noting that increasing the thickness of a dipole antenna increases its bandwidth. Increasing the height also increases the efficiency of the antenna. Increasing the height does induce surface waves that travel within the substrate (which is undesired radiation and may couple to other components).
The following equation roughly describes how the bandwidth scales with these parameters:

$$B \alpha \frac{\varepsilon_r - 1}{\varepsilon_r^2} \frac{W}{L} h$$

*C. Steps involved in designing the patch*
Following steps are taken to design a rectangular micro strip patch antenna.
**Step1:** Calculation of width (W):
The width of the micro strip patch antenna is given by the equation as

$$W = \frac{c}{2f_0 \sqrt{(\varepsilon_r + 1)/2}}$$

W -Width of the patch.
C -Free space velocity of light, 3 x 10! m/s.
**Step 2:** Calculation of the Effective Dielectric Constant.
This is based on the height, dielectric constant of the dielectric and the calculated width of the patch antenna.
*D. Feeding technique*

The feeding technique used here is corporate feeding. Corporate Feed In the corporate feed configuration, the antenna elements are fed by 1:n power divider network with identical path lengths from the feed point to each

$$\varepsilon_{eff} = \frac{\varepsilon_r + 1}{2} + \frac{\varepsilon_r - 1}{2}[1 + 12\frac{h}{W}]^{-\frac{1}{2}}$$

**Step 3:** Calculation of the Effective length

$$L_{eff} = \frac{c}{2f_0 \sqrt{\varepsilon_{eff}}}$$

**Step 4:** Calculation of the length extension ΔL

The following equation gives the length extension as

$$\Delta L = 0.412 h \frac{(\varepsilon_{eff} + 0.3)(\frac{W}{h} + 0.264)}{(\varepsilon_{eff} - 0.258)(\frac{W}{h} + 0.8)}$$

h - Thickness

**Step 5:** Calculation of actual length of the patch element. The advantages of this topology include design simplicity, flexible choice of element spacing, and broader bandwidth, and they are amenable to integration with other devices such as amplifiers and phase shifters. The disadvantage of this type of array is that it requires more space for feed network. For large arrays, the length of feed lines running to all elements is prohibitively long, which results in high insertion loss. The insertion loss is even more pronounced at milimetre-wave frequencies, thereby adversely degrading gain of the array. At higher frequencies, the feed lines laid on the same plane as the patches will also radiate and interfere with the radiation from the patches. Figure 5.1 shows the scheme of corporate feed. It consists of transmission lines, bends, power splitters or T junctions and quarter wave transformers.

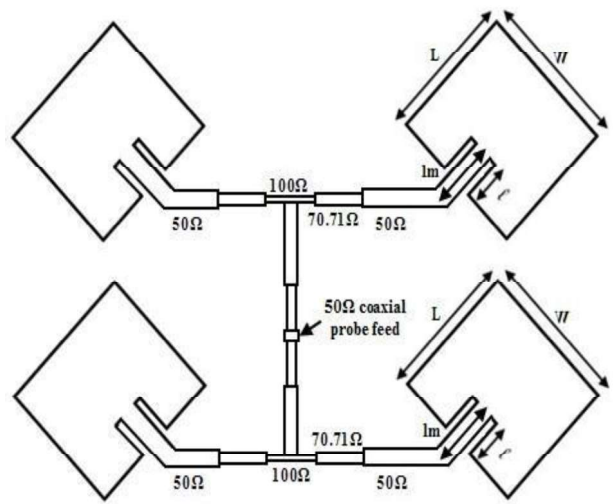

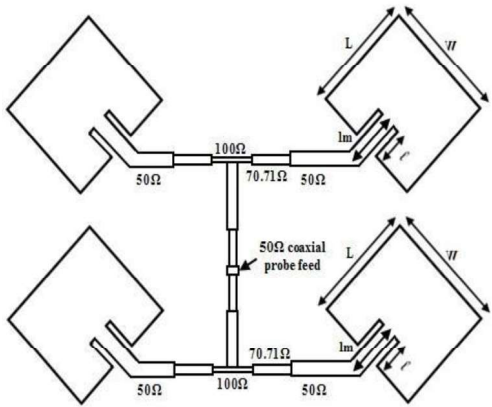

Fig 2.1 corporate feed model

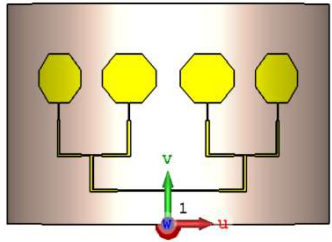

Fig 2.2.a conformal with straight surface

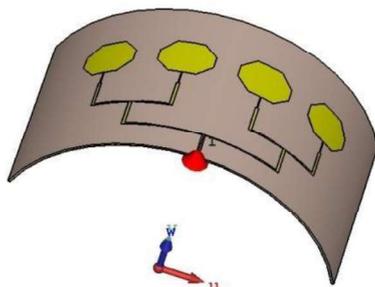

Fig 2.2.b conformal with bend surface.

The above figures shown is the designed model of 1x4 conformal antenna with corporate feed the Layer diagram of the antenna is shown below

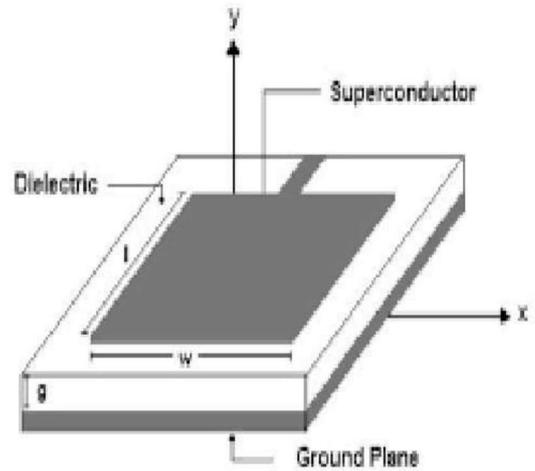

Fig 2.3 layers of microstrip patch antenna.

It consists of three layers namely patch, substrate and ground. The first layer would a super conducting metal like copper with a thickness of about 0.0345mm.we have used an tetrahedral patch for optimization. The second layer will be dielectric material we have used rogers RT6002(lossy) material with permittivity

### III. ANALYSIS

*A. S parameter of the designed antenna*

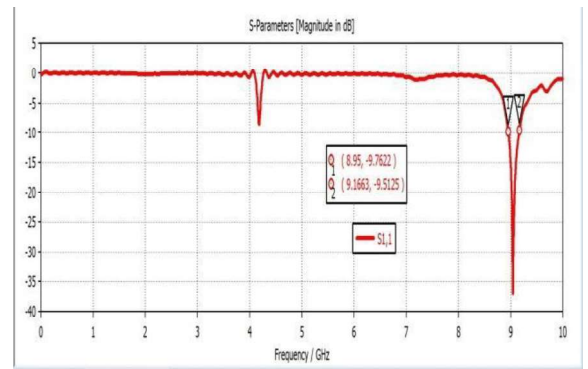

Fig 3.1 S parameter of the radiating antenna.

The above figure shows the s- Parameter of the designed antenna it indicates that it can be operated at frequency of 9GHz with a return loss of about -35.5 db. . Generally bandwidth of the antenna is found from the return loss graph. Bandwidth is obtained from the intersection of return loss graph with -10dB of the graph. The difference between the upper cut off frequency and lower cutoff frequency gives the bandwidth of the antenna.

B. *Gain of the designed antenna*

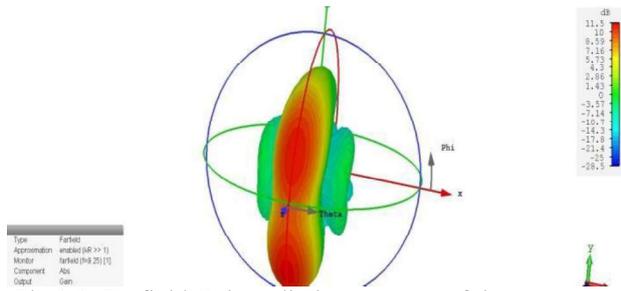

Fig 3.2 Farfield Gain radiation antenna of the antenna.

Gain is a measure of the ability of the antenna to direct the input power into radiation in a particular direction and is measured at the peak radiation intensity. An isotropic antenna radiates equally in all directions. The simulated gain of gain of 11dB is obtained for frequency 9.25GHz.

The Fig 3.3 represents the 2D-directivity of the antenna. Generally 2-D graph represents the polar plot. From the polar plot the magnitude of main lobe and side lobe is obtained. Here the main lobe of magnitude 2.14dB is obtained and the side lobe magnitude of 11.4 dB. Since the main lobe magnitude greater than side lobe level, the antenna has better efficiency.

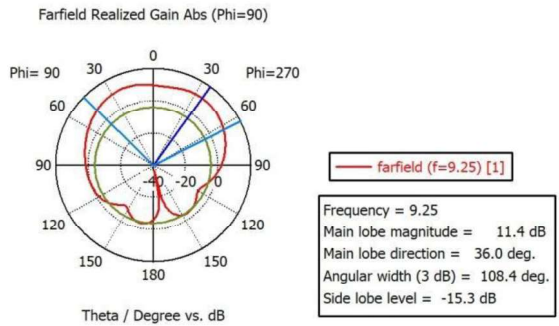

Fig 3.3 2D polar plot of the gain

C. *directivity of the designed antenna*

Directivity measures the power density the antenna radiates in the direction of the strongest emission versus the power density radiated by an ideal isotropic radiator radiating the same total power

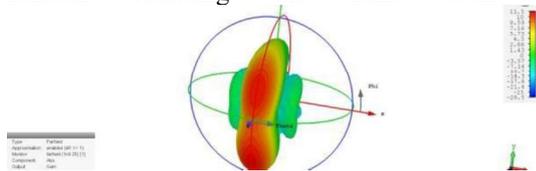

Fig 3.4 Directivity of the antenna

From the Fig 3.4 it is seen that the antenna radiates much along the patch side and radiates less along the downside of the antenna.. The simulated results show a directivity of 12 dBi for frequency 9.25 GHz.

D. *Voltage standing wave ratio of the designed antenna*

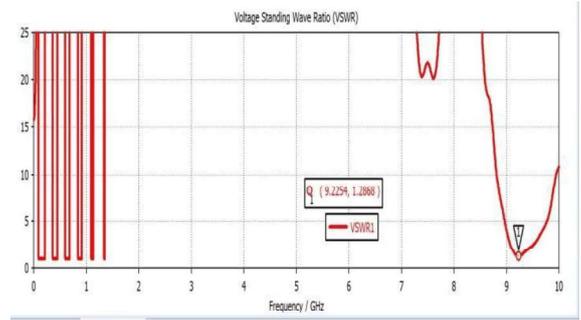

Fig 3.5 Voltage Standing Wave Ratio

From the Fig 6.3 it is seen that the antenna yields minimum
VSWR value of 1.2868 for the frequency

9.225GHz, IV. APPLICATION OF

CONFORMAL ANTENNA

- In avionics applications it is used to transmit and receive information
- In military aircrafts antennas are embedded on the curvy surface to reduce aerodynamic drag.
- Since they cut costs they are used in basic civilian applications like train antennas, car radio antennas and cellular base station to save space and make it less virtually intrusive.
- They have been also used in cars for Anti Brake locking System.(ABS).
- The conformal antenna ahs been also used in medical field for monitoring of body.

V. CONCLUSION AND FUTURE WORKS

This project provides the design and fabrication of conformal antenna which operates at X band. The proposed design of conformal antenna is designed using Computer Simulation Technology (CST) Studio Suite software. Its return loss, radiation pattern, VSWR, directivity, gain is obtained. The proposed antenna can be fabricated using Rogers substrate. It is observed that the results are very good and it can be fabricated.

For the desired application in aircraft. In future, increasing various number of patches in the design can further enhance the results. More gain and directivity can be achieved.